\documentclass[reqno,onecolumn,letterpaper,11pt]{amsart}
\usepackage{amsmath,fullpage,calc,enumerate}
\usepackage{graphicx,psfrag,multicol}
\DeclareGraphicsExtensions{.jpg,.eps,.mps,.png}

\allowdisplaybreaks[1] \numberwithin{equation}{section}
\newtheorem{theorem}{Theorem}[section]

\newtheorem{lemma}[theorem]{Lemma}

\theoremstyle{definition}

{\begin{list}{}{%
   \settowidth{\labelwidth}{\textbf{#1:}}%
   \setlength{\parsep}{0.0cm}%
   \setlength{\itemsep}{0.10cm}%
   \setlength{\leftmargin}{\labelwidth+\labelsep}}}%
{\end{list}}
{\begin{list}{}{%
   \settowidth{\labelwidth}{\textsf{#1)}}%
   \setlength{\parsep}{0.0cm}%
   \setlength{\itemsep}{0.10cm}%
   \setlength{\leftmargin}{\labelwidth+\labelsep}}}%
{\end{list}}
\title{Bottleneck flows in networks}

\author[Abraham P. Punnen Ruonan Zhang ]{Abraham P. Punnen and Ruonan Zhang}

\begin{document}
\thanks {Abraham P. Punnen, Department of Mathematics, Simon Fraser University Surrey,
Central City, 250-13450 102nd AV, Surrey, British Columbia,V3T 0A3,
Canada, Email: {\tt
  apunnen@sfu.ca}}
\thanks {Ruonan Zhang, Department of Mathematics, Simon Fraser University Surrey,
Central City, 250-13450 102nd AV, Surrey, British Columbia,V3T 0A3,
Canada. {\tt rza1@sfu.ca}}

\begin{abstract}
The bottleneck network flow problem (BNFP) is a generalization of
several well-studied bottleneck problems such as the bottleneck
transportation problem (BTP), bottleneck assignment problem (BAP),
bottleneck path problem (BPP), and so on. In this paper we provide a
review of important results on this topic and its various special
cases. We observe that the BNFP can be solved as a sequence of
$O(\log n)$ maximum flow problems. However, special augmenting path
based algorithms for the maximum flow problem can be modified to
obtain algorithms for the BNFP with the property that these
variations and the corresponding maximum flow algorithms have
identical worst case time complexity. On unit capacity network we
show that BNFP can be solved in $O(\min \{{m(n\log
n)}^{\frac{2}{3}}, m^{\frac{3}{2}}\sqrt{\log n}\})$. This improves
the best available algorithm by a factor of $\sqrt{\log n}$. On unit
capacity simple graphs, we show that BNFP can be solved in $O(m
\sqrt {n \log n})$ time. As a consequence we have an $O(m \sqrt {n
\log n})$ algorithm for the BTP with unit arc capacities.
\end{abstract}

\maketitle

\section{Introduction} \vspace{0.4cm}

Let $G(V,E)$ be a directed graph such that ${|V| = n}$ and ${|E| =
m}$. For each arc ${(i,j)\in E}$, a weight $c_{ij}$ and a capacity
$u_{ij}$ are prescribed. Also, for each node $i\in V$ an integer
$b_i$ is associated. Then the \textsl{bottleneck network flow
problem (BNFP)} can be formulated mathematically as follows:

\begin{tabbing}
xxx \= xxx \= xxx \= xxxxx \= xxxxxxxx \= xxxxxxxx \= xxx \= XXX
\=XXX \kill
\>BNFP:\>\>\>\>Minimize ${\max\{c_{ij}: x_{ij} > 0\}}$\\
\>\>\>\>\> Subject to\\
\>\>\>\>\> ${\sum\limits_{\{j:(i,j) \in E\}}{x_{ij}} - \sum
\limits_{\{j:(j,i) \in E\}}{x_{ji}} = b_i \ \ \ \  \forall i \in V}$ \\
\>\>\>\>\> ${0 \leq x_{ij} \leq u_{ij}}\ \ \ \ \forall (i,j) \in
E$
\end{tabbing}
\noindent Here $x_{ij}$ is the flow on arc ${(i,j)}$. We assume that
$c_{ij}$ and $u_{ij}$ are integers for all $(i,j)\in E$ and there
are no multiple arcs in $G$. The integer number ${b_i}$ associated
with node $i$ represents the supply or demand at the node. A node
$i$ is called a \textit{supply node} if ${b_i
> 0}$ and a \textit{demand node} if ${b_i < 0}$. If ${b_i = 0}$, then $i$ is a \textit{transshipment node}.
We assume that ${\sum\limits_{i \in V}{b_i} = 0}$.

An interpretation of the bottleneck network flow problem can be
given as follows: Suppose that perishable goods are shipped from the
supply nodes to the demand nodes. The goods can be stored without
damage at the nodes but are perishable in time on transit. Assume
that $x_{ij}$ is the amount of goods shipped along the arc $(i,j)$
and $c_{ij}$ is the shipment time  along the arc $(i,j)$. Then the
BNFP objective function measures the largest time a shipment in
transit and we want to minimize this time.

To the best of our knowledge, the general form of the bottleneck
network flow problem has not been studied in literature except the a
generalization on algebraic flows~\cite{Brucker1985}. However, many
special cases of this problem are well-studied. One of the most well
studied special cases of BNFP is the \textit{bottleneck
transportation problem (BTP)}. In this case, the graph G is
bipartite with the generic bipartition of V as ${V = V_1 \cup V_2}$,
such that ${i \in V_1}$ implies ${b_i \geq 0}$ and ${i \in V_2}$
implies ${b_i \leq 0}$. Most of the literature on bottleneck
transportation problem assumes that the arcs are without capacities
(i.e. $u_{ij}=\infty $) ~\cite {Burkard1980, Derigs1982, Derigs1979,
Finke1979, Garfinkel1971, Grabowski1991, Hammer1969, Hammer1971,
Hochbaum1999, Isermann1984, Swaroop1977, Malhotra1983, Mironov1996,
Russell1983, Shvartin1975, Szwarc1971, Varadarajan1991} although
some papers admit finite capacities.

In BTP, if ${b_i = 1}$ for ${i \in V_1}$, ${b_i = -1}$ for ${i \in
V_2}$, $|V_1|=|V_2|$ and ${u_{ij} = 1}$ for all ${(i,j) \in E}$,
then the BTP reduces to the well known \textit{bottleneck assignment
problem (BAP)} ~\cite {Efrat2001, Armstrong1992, Borodkin1974,
Cechl1995,Derigs1978,  Efrat2000, Eiselt1984, Eley2003,
Garfinkel1971a,Gross1959, Pferschy1995a, Pferschy1996, Punnen1994,
R.2005, Slominski1979, Tarjan1988}.

Another well studied special case of BNFP is the bottleneck path
problem(BPP). Let $s$ and $t$ be two specified vertices in $G$. If
${b(s)=1, b(t)=-1, b_i=0}$ for $i \in V \setminus \{s, t\}$ and
$u_{ij}=1 $ for every ${(i,j) \in E}$, the resulting bottleneck
network flow problem is called \textit{bottleneck path problem
(BPP)} ~\cite {Gordeev1988, Kaibel2006, R.2005, Listrovio1998,
Monnot2003, Punnen1991, Punnen1996, Garfinkel1998}. Further, BNFP is
a special case of the bottleneck linear programming problem (BLP)
~\cite {Frieze1975, Seshan1982, Vejmola1981}.

The paper is organized as follows. In section 2 we provide a
literature survey on existing algorithms for various special cases
of BNFP. In section 3 we discuss some basic algorithms for solving
the BNFP. We first observe that BNFP can be solved as an $O(\log n)$
sequence of maximum flow problems. We then identify special maximum
flow algorithms that can be modified to solve BNFP with the same
worst complexity as that of solving just one maximum flow problem by
these algorithms. Section 4 deals with the unit capacity networks.
We first observe that the best known maximum flow algorithm for unit
capacity networks can be extended to handle arbitrary capacities on
arcs incident on source and sink nodes. We then show that BNFP on
unit capacity graphs can be solved in ${O(\min\{(n \cdot \log
n)^{\frac{2}{3}} \cdot m,\ m^{\frac{3}{2}}\cdot \sqrt{\log n}\})}$.
This improves the best known algorithm for BNFP on unit capacity
networks by a factor of $\sqrt{\log n}$. For unit capacity simple
networks, we obtain a complexity bound ${O(m \sqrt {n \log n})}$.
This algorithm can be viewed as a generalization of the algorithm of
Gabow and Tarjan for the BAP~\cite{Tarjan1988} and also provides an
${O(m \sqrt {n \log n})}$ time bound for the bottleneck
transportation problem with unit capacities.

\section{Literature Review}

In this section we provide a review of known results on BTP, BAP and
BPP which are special cases of BNFP. To the best of our knowledge no
review papers on the topic is available. We keep the review section
brief, highlighting only important results. For details, the reader
is referred to the original papers. Let us first consider the BTP.
Most of the known algorithms for this problem can  generally be
categorized into three groups: (1) primal algorithms (2) augmenting
path algorithms and (3) threshold algorithm. Primal algorithms start
with a feasible solution and try to find a better solution. Since
the different objective function values of BTP solutions are at most
${m = |V_1|\times |V_2|}$, the number of improvement steps is
${O(m)}$. Algorithms discussed by Hammer ~\cite {Hammer1969,
Hammer1971}, Garfinkel and Rao ~\cite {Garfinkel1971}, Bhatia,
Swaroop and Puri ~\cite {Swaroop1977} etc. falls in this category.
Another class of algorithms build a solution by means of augmenting
paths. Algorithm proposed by Derigs and Zimmermann ~\cite
{Derigs1979} is an example of such an algorithm. The algorithm
augments flows along bottleneck paths until a feasible(and hence
optimal) solution is obtained. The complexity of this algorithm can
be verified to be ${O(S \cdot f(m, n))}$ where ${S = \sum_{i \in
V_1}{|b_i|}}$ and ${f(m, n)}$ is the complexity of BPP. The
threshold algorithm sets a threshold for the optimal objective
function value and checks the existence of a solution satisfying
this threshold value. Depending on the outcome, the threshold value
is adjusted and the process is continued. Some of the primal
algorithms can also be viewed as a threshold algorithm. For a
discussion on threshold algorithms for general combinatorial
bottleneck problems we refer to the paper by Edmonds and
Fulkerson~\cite{Edmonds1970}. When the number of supply (demand)
nodes are fixed, say $k$, Hochbaum and Woeginger ~\cite
{Hochbaum1999} showed that the BTP can be solved in ${O(n)}$ time. A
Special case of this problem when ${k = 2}$ has been studied by
Varadarajan ~\cite {Varadarajan1991} who gave an ${O(n)}$ algorithm
and Szwarc ~\cite {Szwarc1971}who proposed an ${O(n \cdot \log n)}$
algorithm. Many of the works on BTP are relatively old and these
papers do not discuss complexity results. It is easy to obtain a
binary search version of the threshold algorithm to solve BTP as a
sequence of $O(\log n)$ maximum flows in a bipartite graph. This
result extends in a straightforward way to obtain a threshold
algorithm for the BNFP which also solves $O(\log n)$ maximum flows.

Most of the algorithms known for the bottleneck assignment problem
(BAP) can also be categorized as primal algorithms ~\cite
{Gross1959}, augmenting path algorithms ~\cite {Derigs1978} and
threshold algorithms ~\cite {Garfinkel1971a}. These algorithms can
be viewed as specializations of the corresponding algorithms for
BTP. The best known algorithm for BAP is a hybrid algorithm that
uses a binary search based threshold algorithm together with an
augmenting path algorithm. Using this approach, Garbow and Tarjan
~\cite {Tarjan1988} obtained an algorithm of complexity ${O(m \sqrt
{n \log n})}$ to solve BAP. In the threshold phase of this
algorithm, a ``relaxed" feasibility problem is considered to obtain
a partial solution, which is extended into a full solution by means
of augmenting paths. It is the best known time bound for BAP on
sparse graphs. Using a similar approach, Punnen and Nair ~\cite
{Punnen1994} proposed an ${O(n\sqrt{nm})}$ algorithm by considering
a slightly different ``relaxed" problem. This bound is the best
known for solving BAP on dense graphs. When the arc weights are
Euclidian distances, Efrat, Itai and Katz ~\cite {Efrat2001}
proposed an ${O(m)}$ algorithm. When $V \subset {\mathbb{R}}^d$,
Efrat and Katz ~\cite {Efrat2000} proposed an $O(n^{1.5})$ time
algorithm for $d\leq 6$, and a subquadratic time algorithm for
$d>6$. If the underlying norm is $L_\infty$, then the complexity
bound is $O(n^{1.5}{\log}^{0.5} n)$ for $d>2$. When ${c_{ij} = a_i
\cdot b_j}$, Eiselt and Gerchak ~\cite {Eiselt1984} proposed a
simple non-iterative scheme. Probabilistic results on BAP are
discussed by Pferschy~\cite{Pferschy1995a} and specially structured
cost matrices are considered by Cechl\'{a}rov\'{a}~\cite{Cechl1995},
Eiselt and Gerchak~\cite{Eiselt1984}.

A natural approach to solve the bottleneck path problem (BPP) is to
consider modifications of the shortest path algorithms. Many authors
considered modifications of the Dijkstra's algorithm for shortest
path ~\cite {Dijkstra1959} to solve BPP ~\cite {Listrovio1998}. The
complexity of such an algorithm is $O(n^2)$ for a straightforward
implementation. Using Fibonacci heaps, the method can be implemented
in $O(m+n\log n)$ time ~\cite {Ahuja1993}. Listrovio and Khrin
~\cite {Listrovio1998} also proposed a related algorithm explained
using $s-t$ cuts. Their algorithm starts from an $s-t$ cut $K=[S,\
\bar{S}]$, where $S=\{s\},\ \bar{S}=V-\{s\}$. The maximum capacity
of this cut is set to be a lower bound of the objective function
value, and $K$ is iteratively modified by increasing $S$ and
decreasing $\bar{S}$ until the sink node $t \in S$. Fernandez,
Garfinkel and Arbiol~\cite{Garfinkel1998} presented a binary search
based threshold algorithm ~\cite {Garfinkel1998} for BPP. This paper
also discusses an application of BPP in the context of combining
(mosaicking) two or more aerial photographs into a single image in
the production of photographic maps. Inspired by an algorithm of
Gabow and Tarjan for bottleneck arborescence problem, Punnen ~\cite
{Punnen1996} showed that if a bottleneck combinatorial optimization
problem of size $m$ with ordered weights can be solved in
$O(\xi(m))$ time, then the problem with arbitrary weights can be
solved in $O(\xi(m)\log^*(m))$ time, where $\log^*n$ is the iterated
logarithm of $m$. As a consequence, the BPP can be solved in
$O(m\log ^*m)$ time. Combining this approach with modification of
Dijkstra's algorithm discussed earlier, which uses Fibonacci heaps,
the best known complexity for BPP on a directed graph is
$O(\min\{m+n\log n, m\log ^*m\})$. Georgiadis~\cite{Georgiadis2003}
showed that BPP can be solved in $O(T(m))$ time where $T(m)$ is the
time for sorting the edge costs of the underlying graph. BPP on an
undirected graph is simpler and can be solved in linear time using a
binary search based threshold algorithm coupled with subgraph
contractions. For details of this algorithm, we refer to
Punnen~\cite {Punnen1991}. Sensitivity analysis for BPP have been
investigated by Ramaswamy, Orlin and Chakravarty~\cite{R.2005}. It
is easy to show that all pair bottleneck path problem on an
undirected graph can be obtained by computing just one minimum
spanning tree.

\section{Basic algorithms for BNFP}

The BNFP can be formulated as a minimum cost flow problem with
exponentially large arc costs~\cite{Gupta1982}. However, this is not
a practical approach to solve the problem. We now consider some
basic algorithms to solve BNFP, which are generalizations of the
corresponding algorithms for the bottleneck transportation problem
(BTP).

For any real number $\alpha$, let ${G(\alpha) = (V_\alpha,
E_\alpha)}$ denote the spanning subgraph of $G$ with ${V_\alpha =
V}$ and ${E_\alpha = \{ (i,j) \in E: c_{ij} \leq \alpha\}}$. Let
${c_{\sigma(1)} < c_{\sigma(2)} < \cdots < c_{\sigma(\varphi)}}$ be
an ascending arrangement of all distinct arc weights of $G$. Let
${\delta = \sum\limits_{i \in S}b_i = \sum\limits_{i \in T}{-b_i}}$,
where $S$ is the collection of supply nodes and $T$ is the
collection of demand nodes.

\textit{The auxiliary graph} ${G^*=(V^*, E^*)}$ corresponding to any
graph $G(V,E)$ is defined as ${V^*= V \cup\{s,t\}}$, ${E^*= E \cup
\{(s,i): i \in S\} \cup \{(j,t):j \in T\}}$, and $s\notin V$,  $
t\notin V$. Here $s$ is called a \textit{source node} and $t$ is
called a \textit{sink node} in $G^*$. The capacity $u_{si}$ of arc
$(s,i)$ is $b_i$ for all $i \in S$ and the capacity $u_{jt}$ of arc
$(j,t)$ is $-b_j$ for all $j\in T$. The weights $c_{si}$ of arc
$(s,i)$ for all $i\in S$ and $c_{jt}$ of $(j,t)$ for all $j\in T$
are set to be 0. Clearly BNFP is feasible if and only if $G^*$ has
an $s-t$ flow of value $\delta$.

\begin{theorem}\label{th1}
Assume that the BNFP is feasible. Let $k \in \{1,2,\ldots
,\varphi\}$ be the smallest index such that the maximum flow in
${G^*(c_{\sigma(k)})}$ is $\delta$. Then any flow $x^0$ in
${G^*(c_{\sigma(k)})}$ of value $\delta$ provides an optimal
solution $\bar{x}$ to BNFP.
\end{theorem}

The straightforward proof of Theorem~\ref{th1} is omitted. Note that
the solution $\bar{x}$ in the above theorem is obtained by simply
dropping the flow values on arcs incident on $s$ and $t$ from $x^0$.

Based on Theorem~\ref{th1} we see that the if we get the value of
$k$, then we can get an optimal solution for BNFP. In fact, $k$ can
be identified by using different search strategies. Using binary
search over the set $\{c_{\sigma(1)}, c_{\sigma(2)}, \ldots ,
c_{\sigma(\varphi)}\}$, by Theorem~\ref{th1}, it can be verified
that BNFP can be solved by solving $O(\log \varphi) = O(\log n)$
maximum flows. We call this algorithm the \textit{binary search
threshold algorithm}. This observation raises an interesting
question: ``Is it possible to solve BNFP using less than $O(\log n)$
maximum flow computations?'' We do not have an answer to this. Later
in section 4 we will see that for unit capacity networks, we can
solve BNFP using $O(\sqrt{\log n})$ approximate maximum flow
computations. For the general BNFP, let us consider a closely
related question: ``Is it possible to modify a maximum flow
algorithm to solve the BNFP within the same time bound as that of
solving the maximum flow problem with the original algorithm?'' As
we show below, this is doable in some cases and the question is open
for other cases.

Perhaps, the simplest such example is the generic augmenting path
algorithm for maximum flows~\cite{Ahuja1993}. Here, we start with
the graph ${G^*(c_{\sigma(1)})}$ and augment flows from $s$ to $t$
by augmenting paths. If a flow value of $\delta$ is reached, we have
an optimal solution to BNFP. Otherwise we add to
${G^*(c_{\sigma(1)})}$ all arcs of weight $c_{\sigma(2)}$ to obtain
the graph ${G^*(c_{\sigma(2)})}$ and search for augmenting path is
continued. Continuing this process by adding new classes of arcs in
the increasing order of weights and the algorithm terminates when a
flow $x^0$ of value $\delta$ is identified. By Theorem 3.1, an
optimal solution to BNFP can be recovered from $x^0$. As in the case
of the generic augmenting path algorithm for maximum flows, the
complexity of this algorithm is $O(m\delta + \varphi\log \varphi) =
O(mnB)$ where $B=\max\limits_{i\in V}|b_i|.$ This algorithm is a
variation of the augmenting path algorithm of Derigs and
Zimmerman~\cite{Derigs1979} designed for BTP. One major difference
is that we do not use bottleneck path computations to identify
augmenting paths which results in slightly improved complexity for
this more general problem. Further the algorithm provides a natural
linkage with the generic augmenting path algorithm for maximum
flows. We call this algorithm the \textit{BNFP augmenting path
algorithm}.

Let us now consider the maximum capacity augmenting path algorithm
for maximum flows~\cite{Edmonds1972}. This algorithm has polynomial
complexity and we show that the algorithm can be easily modified to
solve BNFP without increasing the complexity bound. Rather than
considering maximum capacity augmentations, we augment flows along
paths with large enough residual capacity to avoid maximum capacity
path computations. This is possible because our target the maximum
flow value $\delta$ is known \textit{a priori}. We call the
resulting algorithm for BNFP the \textit{large capacity augmenting
path algorithm}.

Let ${\bar{G} = (\bar{V}, \bar{E})}$ be a subgraph of $G^*$
containing $s$ and $t$ and ${\bar{G}_r(x)}$ be the residual
graph~\cite{Ahuja1993} with respect to an ${s-t}$ flow $x$ in
$\bar{G}$.

\begin{lemma}
If $\bar{G}$ has a maximum flow from $s$ to $t$ of value $\delta$,
then for any flow $x$ in $\bar{G}$, there exists an augmenting path
in $\bar{G}_r(x)$ with residual capacity at least
${\frac{\delta-v(x)}{m^*}}$, where $m^*=|E^*|$ and $v(x)$ is the
value of flow $x$.
\end{lemma}

\begin{proof}
Since $\bar{G}_r(x)$ is the residual graph of $\bar{G}$ with respect
to flow $x$, then in $\bar{G}_r(x)$, a flow of value $\delta-v(x)$
can be represented as path flows. By the Flows Decomposition Theorem
~\cite {Ahuja1993} (\textit{pp 79-81}), at most $\bar{m}$ s-t paths
have non-zero flow, therefore there must be an augmenting path in
$\bar{G}_r(x)$ with residual capacity at least
${\frac{\delta-v(x)}{\bar{m}}}$. Since $m^*>\bar{m},
\frac{\delta-v(x)}{\bar{m}} \geq \frac{\delta-v(x)}{m^*}$ and the
result follows.
\end{proof}

Lemma 3.2 implies that we can modify the augmenting path algorithm
by performing each augmentation along a path with capacity at least
$\frac{\delta-v(x)}{m^*}$. If no such path exits, we can safely
conclude that $\bar{G}$ does not have a maximum $s-t$ flow  of value
$\delta$. Based on this idea, we present the large capacity
augmenting path algorithm below:

\begin{tabbing}
xxx \= xxx  \=xxxx \=xxxx \=xxxxx \=xxxxx \=xxxx \=XXX \kill
\textbf{Algorithm} \textit{large capacity augmenting
path}\\\textbf{begin}\\\> construct $G^*$ from $G$ and solve a
maximum flow problem on $G^*$;\\\> \textbf{if} the maximum flow
value $<\delta$ \textbf{then stop.} BNFP is infeasible;\\\>
\textbf{else do}\\\>\>let ${c_{\sigma(1)} < c_{\sigma(2)} < \cdots
< c_{\sigma(\varphi)}}$ be an ascending arrangement of all\\
\hspace{2.5cm} distinct arc weights of $G$;\\\>\>${k = 0,\ x^0 =}$
zero flow, ${v(x^0) = 0}$;
\\\>\>\textbf{repeat}\\\>\>\>${k = k + 1}$;\\\>\>\> let $x^k:=x^{k-1},\
v(x^k):=v(x^{k-1})$;\\\>\>\>\textbf{begin}\\
\>\>\>\>obtain the residual graph $G^*_r(c_{\sigma(k)})$ with
respect to the \\ \hspace{4.0cm}flow $x^{k-1}$ from
$G^*_r(c_{\sigma(k-1)})$
by adding the arcs $(i,j)$ \\ \hspace{4.0cm}whose weight 
$c_{ij}=c_{\sigma(k)}$;\\\>\>\>\> \textbf{while}
($G^*_r(c_{\sigma(k)})$ contains an augmenting path $P$ of\\
\hspace{4.0cm} residual capacity at least $\frac{\delta -
v(x^{k-1})}{m^*}$) \textbf{do}
\\\>\>\>\>\>let
$\varepsilon$ be the residual capacity of $P$;\\\>\>\>\>\> augment
$\varepsilon$ units of flow along $P$;\\ \>\>\>\>\> $v(x^k)=v(x^k)
+\varepsilon$;\\\>\>\>\>\> update $x^k$ and $G^*_r(c_{\sigma(k)})$;\\
 \>\>\>\> \textbf{end while};\\\>\>\>\textbf{end if};\\
\>\> \textbf{until} ${(v(x^k)= \delta)}$\\\>\>compute and output
the optimal BNFP solution corresponding to $x^k$;\\\>
\textbf{end};\\\textbf{end};
\end{tabbing}

To test the conditions of the while loop of the above algorithm, we
can construct a graph $\hat{G}^*_r(c_{\sigma(k)})$ by removing all
the arcs in $G^*_r(c_{\sigma(k)})$ whose residual capacity is less
than $\frac{\delta - v(x^{k-1})}{m^*}$. Then $G^*_r(c_{\sigma(k)})$
has an augmenting path of capacity at least $\frac{\delta -
v(x^{k-1})}{m^*}$ if and only if $\hat{G}^*_r(c_{\sigma(k)})$ has an
$s-t$ path.

To establish the complexity of the large capacity augmenting path
algorithm, we prove the following theorem, which is a variation of a
result by Edmonds and Karp ~\cite {Edmonds1972} and Goldfarb and
Chen ~\cite {Goldfarb1997} for the maximum capacity augmenting path
algorithm for the maximum flow problem.

\begin{theorem}
In the large capacity augmenting path algorithm, the number of
augmentations is $O(m \log \delta) = O(m \log (nB))$, where ${B=
\max\limits_{i\in V}|b_i|}$.
\end{theorem}

\begin{proof}
Let $v(x^1),\ v(x^2),\cdots, v(x^k)=\delta$ be a sequence of flow
values generated by the large capacity augmenting path algorithm.
Thus $k$ is the total number of augmentations performed. Assume
${d_i = v(x^{i+1}) - v(x^i)}$ and ${\Delta_i = \delta - v(x^i)}$,
then\\ \\\vspace{0 cm} \hspace{2.3cm} $d_i = v(x^{i+1}) - v(x^i) = \Delta_i - \Delta_{i+1} \ \ \ \ \ \ \ \  (1)$\\
Since the augmenting paths have capacity at least
${\frac{\delta-v(x^i)}{m^*}}$, we have\\
\vspace{0 cm} \hspace{3.0cm} ${d_i \geq \frac{\delta - v(x^i)}{m^*}}$,\\
\vspace{0 cm} \hspace{2.0cm} and hence \ ${\Delta_i \leq
m^*d_i\ \ \ \ \ \ \ \ \ \ \ \ \ \ \ \ \ \ (2)}$\\
From (1) and (2),\\
\vspace{0 cm} \hspace{2.4cm}
$\ \Delta_{i+1} \leq \Delta_i(1- \frac{1}{m^*})$.\\
Therefore \\
\vspace{0 cm} \hspace{3.0cm} $\Delta_p \leq
\Delta_1(1-\frac{1}{m^*})^p\leq \delta(1 - \frac{1}{m^*})^p\leq
\delta e^{- p/m^*}$\\
We want the largest $p$ such that \\
\vspace{0 cm} \hspace{3.2cm}${\Delta_p \geq 1}$. Thus  $ p \leq m \log \delta $\\
Since ${v(x^k) = \delta}$, we have ${k \leq p+1}$. Therefore $k =
O(m^* \log \delta)=\ O(m \log(nB))$.
\end{proof}

\begin{theorem} The large capacity augmenting path
algorithm correctly solves BNFP in ${O(m^2\log (nB))}$ time.
\end{theorem}
\begin{proof}
Starting with a zero flow in $G^*(c_{\sigma(1)})$, the algorithm
looks for the smallest index $k$ such that $G^*(c_{\sigma(k)})$ has
a flow of value $\delta$. At a typical iteration we have a flow
$x^r$ in $G^*(c_{\sigma(r)})$ for some $r$ with value $v(x^r)$ and
$v(x^r)< \delta$. Then introduce arcs $c_{\sigma(r+1)}$ to obtain
the graph $G^*(c_{\sigma(r+1)})$. Clearly $x^r$ is a feasible flow
in $G^*(c_{\sigma(r+1)})$. If the residual graph of
$G^*_r(c_{\sigma(r+1)})$ does not have an augmenting path of value
at least ${\frac{\delta-v(x^r)}{m^*}}$, by Lemma 3.2,
$G^*(c_{\sigma(r+1)})$ does not contain a flow of value $\delta$ and
the arc ${c_{\sigma(r+2)}}$ is added to $G^*(c_{\sigma(r+1)})$ to
obtain $G^*(c_{\sigma(r+2)})$ and ${c_{\sigma(r+2)}}$ becomes a new
lower bound for the optimal objective function value of BNFP.
Otherwise the flow is improved by using an augmenting path of
capacity at least $ \frac{\delta-v(x^r)}{m^*}$. By Theorem 3.3 the
number of augmentations is bounded by ${O(m\log (nB))}$. The
complexity of performing the augmentation step is ${O(m)}$, so the
overall complexity of the algorithm is ${O(m^2\log (nB))}$.
\end{proof}

It may be noted that we could not obtain a variation of the shortest
augmenting path algorithm for maximum flows that solves BNFP with in
the same time bound as the corresponding maximum flow algorithm.

\section{BNFP in Unit Capacity Networks}

On a unit capacity graph, it is well known that the maximum flow
problem can be solved in
$O(\min\{m^{3/2},n^{2/3}m\})$ time ~\cite{Even1975}. Suppose $G$ is a
unit capacity graph on which a BNFP is defined. Then the
corresponding auxiliary graph ${G^*=(V^*, E^*)}$ will be of unit
capacity except for the arcs incident on the source node $s$ and
sink node $t$. We first observe that the maximum flow problem in
such a graph can also be solved in $O(\min\{m^{3/2},n^{2/3}m\})$
time.

 A graph $G$ with a source node $s$ and a sink node $t$
  is called an \textit{almost unit capacity graph} if (1) arcs incident on $s$
or $t$ or both have arbitrary capacities (2) all other arcs are of
unit capacity and (3) any $s-t$ path in $G$ contains at least one
arc which does not incident on $s$ or $t$. The maximum flow problem
restricted to an almost unit capacity graph is called \textit{almost
unit capacity maximum flow problem (AMFP)}. An almost unit capacity
graph $G$ is  simple  if every node in $G$ has at most one incoming
arc or at most one outgoing arc. The corresponding maximum flow
problem is called \textit{almost unit capacity  simple maximum flow
problem (ASMFP)}.

\subsection{Flows in Almost Unit Capacity Graphs}

Let us now discuss the maximum flow problem in almost unit capacity
graphs and almost unit capacity simple graphs. Our algorithms are
similar to the unit capacity maximum flow algorithm of Edmonds and
Karp~\cite{Edmonds1972} as discussed in Ahuja and Orlin \cite
{Ahuja1993}. We only discuss the primary results without proof. An
interested reader could construct the proofs with appropriate
modifications of the corresponding unit capacity case or can find it
in the thesis~\cite{Zhang2005} where details of the algorithms of
this paper are given.

It is well known that the shortest augmenting path
algorithm~\cite{Ahuja1993} solves the maximum flow problem in
$O(mn^2)$ time. On unit capacity graphs, the complexity can be
reduced to $O(mn)$~\cite{Ahuja1993}. The shortest augmenting path
algorithm maintains distance labels that are non-decreasing and
terminates when the distance label $d(s)$ of node $s$ satisfies
$d(s) \geq n$. Let $D \leq n$ be a parameter. In the shortest
augmenting path algorithm, if we discard all nodes $i$ with distance
label $d(i) \geq D$ from further consideration, we get an
approximate version of the shortest augmenting path algorithm. We
refer to this algorithm the \textit{D-shortest augmenting path
algorithm}.

\begin{theorem}\label{thu1} In an almost unit capacity graph, the $D$-shortest
augmenting path algorithm terminates in $O(Dm)$ time.
\end{theorem}
The proof of this theorem can be constructed from similar results
for the unit capacity case and hence omitted. The following theorem
provides an approximation bound for the solution produced by the
$D$-shortest augmenting path algorithm.

\begin{theorem}\label{thu2} Let $G(V,E)$ be an almost unit capacity graph with no
parallel arcs. Suppose $x$ is a flow generated by the $D$-shortest
augmenting path algorithm and $x^*$ be a maximum $s-t$ flow in $G$.
Then (i) $v(x^*)-v(x) \leq \frac{|E|}{D-2}.$ (ii) $v(x^*)-v(x) \leq
(\frac{2|V|}{D-2})^2.$ (iii) If $G$ is a simple almost unit capacity
graph, then $v(x^*)-v(x) \leq \frac{|V|}{D-2}.$
\end{theorem}

Again the proof of this theorem can be constructed by modifying
arguments in the proof of corresponding results for the unit
capacity maximum flow algorithms. Detailed proof is available in the
thesis~\cite{Zhang2005}.

Let $x^0$ be a flow produced by the $D$-shortest path algorithm in
$G$. Extend $x^0$ into a maximum flow in $G$ using the labeling
algorithm~\cite{Ahuja1993}. Note that the labeling algorithm
performs at most $v(x^*)-v(x^0)$ flow augmentations where $x^*$ is a
maximum flow in $G$. Thus this labeling phase can be implemented in
$O((v(x^*)-v(x^0))m)$ time. Combining this with theorems \ref{thu1}
and \ref{thu2}(i), we get a complexity bound of
$O(\frac{m^2}{D}+Dm)$. Choosing $ D = \lceil \sqrt{m}\rceil$, we get
a bound of $O(m^{3/2})$. Likewise, Combining with theorems
\ref{thu1} and \ref{thu2}(ii), we get a complexity bound of
$O(\frac{n^2m}{D^2}+Dm)$. Choosing $ D = \lceil n^{2/3} \rceil$, we
get a complexity bound of $O(n^{2/3}m)$. For simple graphs, we get a
complexity bound on $O(\frac{nm}{D}+ Dm)$. Choosing $D = \sqrt{n}$
we get a complexity bound of $O(m\sqrt{n})$. Summarizing the
forgoing discussion,

\begin{theorem}\label{thu3} The maximum flow problem in almost unit capacity
network can be solved in $O(\min\{m^{3/2},n^{2/3}m\})$ time. For an
almost unit capacity simple graph, the problem can be solved in
$O(m\sqrt{n})$ time.
\end{theorem}

Note that Theorem \ref{thu3} generalizes corresponding results on
unit capacity networks to almost unit capacity graphs. The
discussions of this section are crucial to our improved algorithm
for BNFP on unit capacity networks.

\subsection{Algorithm for BNFP in unit capacity networks}

Let $G(V,E)$ be a unit capacity graph on which a BNFP is defined.
Then, as noted earlier, its auxiliary graph $G^*(V^*,E^*)$ is an
almost unit capacity graph. Thus combining Theorem \ref{thu3} with
the binary search threshold algorithm for BNFP discussed in Section
3, it can be seen that BNFP on unit capacity graphs can be solved in
$O(\min\{m^{3/2},n^{2/3}m\}\log n)$ time  and in $O(m\log
n\sqrt{n})$ time on unit capacity simple graphs. We now show that we
can improve these bounds by a factor of $O(\sqrt{\log n})$.

Our algorithm first computes a lower bound on the optimal objective
function value of BNFP using an approximate version of the binary
search threshold algorithm. This also generates a flow in $G^*$
which is a partial solution to BNFP. This partial solution is then
extended to a solution to BNFP using our BNFP augmenting path
algorithm.

Let ${c_{\sigma(1)} < c_{\sigma(2)} < \cdots < c_{\sigma(\varphi)}}$
be an ascending arrangement of all distinct arc weights of $G$. Note
that the optimal objective function value of BNFP is one of these
$c_j$ values. Consider the graph $G^*(c_{\sigma(k)})$ for some $k$
with edge set, say $E^k$. Let $x^k$ be the flow produced by the
$D$-shortest augmenting path algorithm on $G^*(c_{\sigma(k)})$. If
$\delta - v(x^k) > \frac{|E^k|}{D-2}$ then by Theorem \ref{thu2}
(i), we can conclude that the maximum flow in $G^*(c_{\sigma(k)})$
is strictly less than $\delta$ and hence $c_{\sigma(k)}$ is a lower
bound. If $\delta - v(x^k) \leq \frac{|E^k|}{D-2}$, then the maximum
flow value in $G^*(c_{\sigma(k)})$ may or may not be equal to
$\delta$. In this case, we make a heuristic decision to set
$c_{\sigma(k)}$ as an approximate upper bound on the optimal
objective function value. Using this search strategy we present our
\textit{approximate binary search threshold algorithm} (Algorithm
ABST) below. Without loss of generality assume $G^*$ contains a
maximum flow of value $\delta$.

\begin{tabbing}
xxx \= xxx  \=xxxx \=xxxx \=XXX \kill \textbf{Algorithm}
\textit{ABST}\\\textbf{begin}\\\> construct $G^*$ from $G$ \\\>\>let
${c_{\sigma(1)} < c_{\sigma(2)} < \cdots < c_{\sigma(\varphi)}}$ be
an ascending arrangement of all \\ \vspace{0 cm} \hspace{2.0 cm}
distinct arc weights of $G$;\\\>\>let ${l = 1,\ u= \varphi}$;\\
\>\>\textbf{while} ($u-l \geq 1$) \textbf{do}
\\ \>\>\> set k:${= \lfloor \frac{l + u}{2}\rfloor}$;\\
\>\>\>construct $G^*(c_{\sigma(k)})$\\\>\>\>let $x^k_{D+2}$ be the
flow produced by the $(D+2)$-shortest augmenting \\
\hspace{3.0 cm}  path algorithm on $G^*(c_{\sigma(k)})= (V^*,
E^*)$
\\ \>\>\> \textbf{if} $\delta-v(x^k_{D+2})\leq \frac{|E^*|}{D}$ \textbf{then} $u=k$;\\
\>\>\> \textbf{else} $l=k+1$;\\
\>\> \textbf{end while};\\\>\>(comment: at this stage $u=l $.)
\\\>\> output the flow $\bar{x}$ produced by the $D+2$ shortest
augmenting path algorithm on $G^*(c_{\sigma(u)})$. \\\textbf{end};
\end{tabbing} \vspace{0.35cm}

\begin{theorem}
Let $\bar{x}$ be the flow produced by Algorithm ABST and
$\bar{c}=\max\{c_{ij}:\bar{x}_{ij}>0\}= c(\sigma_p)$. Then $\bar{c}$
is a lower bound for the optimal objective function value of BNFP.
Further, $\delta-v(\bar{x})\leq \frac{2|E|}{D}$.
\end{theorem}

\begin{proof}
Obviously the starting lower bound $c_{\sigma(1)} \leq c^*$. The
index $l$ of the lower bound value is updated only when we are
guaranteed that $G^*(c_{\sigma(l)})$ contains no flow of value
$\delta$ and hence $c_{\sigma(l+1)} \leq c^*$. Thus by Theorem 3.1,
$\bar{c}\leq c^*$. From the algorithm it can be easily verified that
$\delta-v(\bar{x})\leq \frac{|E^*|}{D}\leq \frac{2|E|}{D}$.
\end{proof}

The complexity of Algorithm ABST is $O(Dm\log n)$. If $v(\bar{x}) =
\delta$ then an optimal solution to BNFP can be constructed from
$\bar{x}$. Otherwise we perform the following \textit{extension
phase}. Starting with $\bar{x}$ and $G^*(\bar{c})$, the BNFP
augmenting path algorithm can be used to compute an optimal solution
to BNFP by performing at most $O(\frac{m}{D})$ flow augmentations.
This can be done in $O(\frac{m^2}{D}) $ time. Thus BNFP can be
solved in $O(Dm\log n + \frac{m^2}{D})$ time. Choosing $D =
\sqrt{\frac{m}{\log n}}$ we get the complexity bound
$O(m^{\frac{3}{2}}\sqrt{\log n})$ to solve BNFP on unit capacity
graphs.

A similar algorithm can be obtained if we replace the condition
$\delta-v(x)\leq \frac{|E^*|}{D}$ by $\delta-v(x) \leq
{(\frac{2|V^*|}{D})}^2$ in Algorithm ABST. In this case the number
of flow augmentation steps in the extension phase is at most
${(\frac{4n}{D})}^2$. This leads to a complexity bound of
$O(\frac{n^2}{D^2m}+Dm\log n)$ time. Choosing $D= {(\frac{n^2}{\log
n})}^{\frac{1}{3}}$, we get the bound $O(m{(n\log
n)}^{\frac{2}{3}})$. The foregoing discussions can be summarized as

\begin{theorem}
The BNFP on a unit capacity graph can be solved  in  $O(\min
\{{m(n\log n)}^{\frac{2}{3}}, m^{\frac{3}{2}}\sqrt{\log n}\})$ time.
\end{theorem}

Note that this bound is better than the bound obtained in the
beginning of this section by a factor of $\sqrt{\log n}$.

For the case of the bottleneck flows in unit capacity simple graphs,
we replace the condition ``if $\delta-v(x) \leq \frac{|E^*|}{D}$" by
``if $\delta-v(x) \leq \frac{|V^*|}{D}$", in algorithm ABST and then
as in the previous discussions, at the end of Algorithm ABST the
extension phase performs at most $\frac{|V|}{D}$ flow augmentations.
Since each augmentation takes $O(m)$ time,  it takes
$O(\frac{nm}{D})$ time to complete the extension phase. Thus the
problem can be solved in $O(\frac{nm}{D}+Dm\log n)$ time. Choosing
$D=\sqrt{\frac{n}{\log n}}$, we get the complexity bound
$O(m\sqrt{n\log n})$. Summarizing these discussions,

\begin{theorem}
The BNFP on a unit capacity simple graph can be solved in
$O(m\sqrt{n\log n})$ time.
\end{theorem}

Note that the bottleneck assignment problem and the bottleneck
transportation problem with unit capacities are special cases of the
BNFP on unit capacity simple networks. Thus we have $O(m\sqrt{n\log
n})$ bounds for these problems as well. This algorithm can be viewed
as an extension of the algorithm of Gabow and
Tarjan~\cite{Tarjan1988} for the BAP.

\section{Conclusion}

In this paper we have considered the Bottleneck Network Flow Problem
(BNFP), which is a generalization of several well studied Bottleneck
problems including the Bottleneck Transportation Problem (BTP),
Bottleneck Assignment Problem (BAP) and Bottleneck Path Problem
(BPP). Some basic algorithms have been discussed. It is observed
that BNFP can be solved as an $O(\log n)$ sequence of maximum flow
problems. Special maximum flow algorithms are identified that can
easily be modified to solve BNFP in the same worst case time bound
as that of solving one maximum flow problem by these algorithms. The
class of such maximum flow algorithms include generic augmenting
path algorithms and maximum capacity augmenting path algorithm. We
could not establish a similar property for the shortest augmenting
path algorithm and we have't investigated preflow push algorithms in
this context.

We have also considered a special case where the arc capacities are
unity. It is well known that ~\cite {Even1975} the maximum flow
problem on unit capacity graphs can be solved in
$O(\min\{m^{\frac{3}{2}}, n^{\frac{2}{3}}m\})$ time. We showed that
the same time bound can be achieved for solving the maximum flow
problem on an almost unit capacity graph, where the capacities of
arcs incident on source and sink nodes are allowed to be arbitrary.
This together with the binary search threshold algorithms shows that
BNFP on unit capacity networks can be solved in
$O(\min\{m^{\frac{3}{2}}, n^{\frac{2}{3}}m\}\log n)$ time. We Also
proposed another algorithm to solve the problem with an improved
complexity of $O(\min \{{m(n\log n)}^{\frac{2}{3}},
m^{\frac{3}{2}}\sqrt{\log n}\})$. For the bottleneck flow problems
in unit capacity simple graphs, we proposed an $O(m\sqrt{n\log n})$ algorithm.
As a byproduct, we get an $O(n\sqrt{m\log n})$ algorithm for the
bottleneck transportation problem with unit capacities.

An obvious question is if these algorithms can be improved? For the
general BNFP, it would be interesting to examine what are the
maximum flow algorithms that can be modified to solve BNFP without
increasing the worst case complexity. As noted earlier, BNFP can be
solved as a sequence of $O(\log n)$ maximum flows. Likewise, the
maximum flow problem can be solved as an $O(\log(nU))$ sequence of
the BNFPs, where $U = \max\{u_{ij}: (i,j) \in E\}$. It would be
interesting to investigate further complexity relationships between
these problems.

\vspace{.10cm} \nocite{*}
\bibliographystyle{plain}
\bibliography{Adata}
\end{document}